
%
%

\input manumac

\def\kp{\kappa}
\def\p{{\bf p}}
\def\q{{\bf q}}
\def\k{{\bf k}}
\def\lt{\lambda_T}
\def\om{\omega}
\def\ddp{{d^2\p\over (2 \pi)^2}}
\def\ddq{{d^2\q\over (2 \pi)^2}}
\def\ddk{{d^2\k\over (2 \pi)^2}}
\def\G{{\cal G}^{0}}
\def\x{x^{\ast}}

\twelvepoint
\doublespace
\rightline{EHU-FT-93}
\rightline{April 1993}
\vskip 2 cm

\centerline{ \bf THREE-LOOP CALCULATION OF THE }
 \bigskip
\centerline { \bf ANYONIC FULL CLUSTER EXPANSION   }

\vskip 2 cm
\centerline { R. Emparan $^{a}$}
\medskip
\centerline {and}
\medskip
\centerline { M. A. Valle Basagoiti $^{b}$}
\vskip .5 cm
\centerline {\it $^{a}$ Departamento de F\'\i sica,}
\centerline {\it $^{b}$ Departamento de F\'\i sica Te\'orica,}
\centerline {\it Universidad del Pa\'\i s Vasco, Apartado 644, 48080 Bilbao,
Spain}
\vskip 4 cm
\centerline {\bf Abstract}
\noindent
We calculate the perturbative correction to every cluster coefficient of a gas
of anyons through second order in the anyon coupling constant, as
described by Chern-Simons field theory.

\vfill
\eject

It has been known for some years that two dimensional
physics offers the possibility of fractional statistics
(anyons) interpolating between bosons and fermions [1].
Either for its possible relevance in some areas of
condensed matter physics, or for the deep implications it
has on the concept of identical particles, the study of the
thermodynamical properties of the anyonic system has
deserved great effort since the past decade. However,
until recently, the low density, high temperature expansion could only
provide the second virial coefficient [2]. Some restricted information
about the third virial coefficient was obtained in refs. [3,4]. Numerical
computations have also been performed [5-7]. The root of the problem
is the unsolvability of the $N$-anyon quantum mechanical
system beyond $N=2$.

This obstacle can be overcome making use of second
quantized theory. The price one has to pay is that
perturbative field theory only provides an expansion in
powers of $\theta$ (the statistical parameter) near
bosonic or fermionic statistics.

In this approach, a problem is posed by the singular
nature of the short range statistical interaction. In refs. [8-11]
a solution is proposed introducing a two-body
non-hermitian interaction. Divergences must be carefully controlled with
harmonic regulators or boundary conditions. In this way,
Dasni\`eres de Veigy and Ouvry [11] have been able to compute the partition
function to order $\theta ^2$, expanded to all orders in the fugacity $z=\exp
(\mu /T)$.

Recently another model has been considered to solve the problem [12].
There, fractional  statistics is implemented in standard way through
non-relativistic matter interacting with a Chern-Simons gauge field. UV
divergences are cancelled with a two-body $\delta$-function potential when
the starting statistics is bosonic. Starting from fermionic statistics does
not require such interaction. This model is used for to the computation
of the second virial coefficient, to three-loop order. In this letter we
develop further the model and obtain the solution to all orders in $z$, to the
same loop order. Remarkably, our results are in full agreement with those of
refs. [10,11] \footnote{$^\dagger$}{\tenpoint The relation between both
approaches is discussed in ref. [12].}.
\eject

The setting of our model, as proposed in ref. [12] consists of a spinless
non-relativistic self-coupled matter field $\psi$ interacting with
Chern-Simons gauge bosons. The Lagrangian density
will be given by
$$
{\cal L}(t,{\bf x}) = -{1 \over 2 \kappa}\,\partial_t\,{\bf
a} \times{\bf a} \,+ {1\over\kappa}\, a_0 B \, -
{1\over 2 \rho}\,(\nabla\cdot{\bf a})^2 \, +
\psi^\dagger i D_0 \psi - {1\over 2 m}\mid\! {\bf D}
\psi\!\mid^2 \, +\, \mu\,\psi^\dagger \psi -
{\alpha\over 4}\, (\psi^\dagger \psi)^2,
$$
$$
 D_0 = \partial_t + i\, a_0, \qquad \
{\bf D}=\nabla - i\,{\bf a}\,\,. \eqno(1)
$$
Here $B=\nabla\times {\bf a}$ is the magnetic field, $\mu$ is the chemical
potential and $\rho$ is a gauge fixing parameter. The Coulomb gauge to be used
refers to the choice $\rho=0$. The anyon coupling $\kp$ can be related to the
statistical angle parameter $\theta$ as $\kp=2\theta$. Also,
the strength of the contact interaction $\alpha$ must be taken [12] as
$$
\alpha = (1+\zeta ){\kp\over m}\, , \eqno(2)
$$
($\zeta =1$ for bosons; $\zeta =-1$ for fermions). With this choice, the
divergences appearing for fiducial bosonic statistics are exactly cancelled;
for fermions, $\alpha$ vanishes, not needing additional interactions.

In the imaginary-time formalism of thermal field
theory [13], the functional integral expression for the grand partition
function involves an integration over imaginary time from $0$ to $\beta =
T^{-1}$ ,
$$
Z(\mu,T,V) =
\int_{\rm (anti)periodic} {\cal D}a_0\,{\cal D}{\bf a}\, {\cal
D}\psi^\dagger{\cal D}\psi\, \exp\,\Bigl( \int_0^\beta d\tau \int
d^2\!x\, {\cal L}(t\!=\!-i\tau,{\bf x}) \Bigr)\,. \eqno(3)
$$
Here (anti)periodic means that the integration over fields
is constrained so that $\psi({\bf x},\beta)=\pm \psi({\bf x},0)$ where
the (lower) upper sign refers to (fermions) bosons.

Now we can proceed to expand in a power series in $\kappa$, by using a set
of diagrammatic rules which follows from (1). These are listed in Table 1.
The diagrams describing the perturbative series for $\ln Z$ have the form
of connected closed loops. It should be noted that there will
be a factor of $\beta A$ left over for each graph, corresponding to the
extensivity of  $\ln Z$. This factor cancels out in expressions for the
pressure.

\eject

The contribution to $\ln Z$ up to order $\kp^2$ involves a set of
three-loop diagrams. The non-zero graphs are shown in fig.1
\footnote{$^\dagger$}{\tenpoint Graph (h) was missing in ref.[12];
since its contribution is of order $z^3$ the results of ref.[12] remain
unaffected.}. Note that all graphs with only one gauge boson line vanish due to
the index summation associated with vertices.
In order to keep the correct order of the operators according to their
$\tau$-values, one must insert a factor $e^{i\omega_{n}\eta}$ whenever a
particle line either closes on itself or is joined by the same instantaneous
interaction line. We take $\eta\rightarrow 0^{+}$ at the end of the
calculation.
This implies the following basic frequency sums
$$
\eqalignno{
{1\over\beta}\sum_{n}{e^{i \omega_{n} \eta} \over i\,\omega_{n} + \mu - {\bf
q}^2/(2 m)} &=  -\zeta\, {1\over \exp{\bigl[\beta\,({\bf q}^2/(2 m) -
\mu)\,\bigr ]}  -\zeta} =  -\zeta\,n_{\q}\,, &(4)\cr
{1\over\beta}\sum_{n}{1  \over i\,\omega_{n} + \mu - {\bf q}^2/(2 m)} &=
-{1\over 2} -\zeta\,n_{\q}\,.&(5)\cr}
$$
As a consequence, graphs (c), (g) and (h) are of order $z^3$.

The above explained formalism allows us to derive the perturbative corrections
to the pressure
$$
P = {T\over A}\ln Z .\eqno(6)
$$
To each order in $\kp$ our results will give the pressure as a function of the
fugacity $z=\exp(\mu /T)$. This corresponds to the cluster expansion
$$
{P(\mu,T)\over T} = {1\over \lt^{2}} \sum_{l=1}^\infty b_l\,z^l \, ,
\eqno (7)
$$
($\lt =(2\pi /mT)^{1/2}$ is the thermal wavelength).
However, the statistical dependence is more clearly manifest in the virial
expansion
$$
{P(n,T)\over T}= {1\over \lt^{2}} \sum_{l=1}^\infty a_l\,(n\lt^2)^l \, ,
\eqno (8)
$$
with
$$
n\equiv N/A=z{\partial\over\partial z}\biggl({P\over T}\biggr)\, .
\eqno (9)
$$
Both expansions can be related easily order by order:
$$
\eqalignno{ a_1 &=1\, , \cr
          a_2 &=-{b_2\over b_1^2}\, , \cr
          a_3 &=4\,a_2^2-2\,{b_3\over b_1^3}\, , &(10) \cr
          &\ldots\cr}
$$
We proceed now to calculate the first two orders of the perturbative
expansion in $\kp$.
\bigskip
\noindent{\it First order}
\medskip
As argued before, to this order, diagrams with one gauge boson line vanish.
This leaves only diagram (a). As a result, fermionic statistics receives no
corrections of order $\kp$. The contribution to the pressure can be written
$$
\eqalignno{
P^{(1)} &=
   -{\alpha\over 2} \biggl[{1\over\beta}\sum_{\om_{1}}
\int \ddp\,\G (\om_{1},\p)\;e^{i\om_1\eta}
\biggr ]^2 \cr
&=-{\alpha\over 2}{1\over\lt^4}\ln^2(1-z) \, . &(11)\cr}
$$
This result agrees with that of refs. [9,11]. It is easy to show from
eqs. (10) and (11) that only the second virial coefficient is corrected to this
order. \bigskip
\noindent{\it Second order}
\medskip
Diagrams (b), (e) and (f) are hardest to compute. They are given by
$$
\eqalignno{
P^{b} &=
  {\alpha^2\over 8\beta^3} \sum_{\om_{1}\;\om_2\;\nu}
\int \ddp\ddk\ddq\;\G (\om_{1},\p)\;\G (\om_{1}+\nu,\p+\q) \cr
&\qquad\qquad\qquad\times\G (\om_{2},\k)\;\G (\om_{2}+\nu,\k+\q) \cr
&\cr
&=(1+\zeta){\kp^2\over 2m}\int\ddp\ddk\ddq\;{n_\p n_{\k +\q} +2\zeta
n_\k n_\p
 n_{\k +\q}\over \q\cdot (\p -\k )}\, , &(12)\cr
}
$$
\medskip
$$
\eqalignno{
P^{e} &=
  -\zeta{\kp^2\over 4m^2\beta^3}\sum_{\om_{1}\;\om_2\;\nu}
\int \ddp\ddk\ddq\;\G (\om_{1},\p)\;\G (\om_{1}+\nu,\p+\q)\cr
&\qquad\qquad\qquad\times\G (\om_{2},\k)\;\G (\om_{2}+\nu,\k+\q)
\;{\bigl[ (\p -\k )\times\q \bigr] ^2\over (\p -\k )^2\;\q ^2} \cr
&\cr
&=-\zeta{\kp^2\over 2m}\int\ddp\ddk\ddq\;{n_\p n_{\k +\q} +2\zeta n_\k n_\p
 n_{\k +\q}\over \q\cdot
(\p -\k )}\;{\bigl[ (\p -\k )\times\q \bigr] ^2\over (\p -\k )^2\;\q ^2}
\, ,&(13)\cr }
$$
\medskip
$$
\eqalignno{
P^{f} &=
  \zeta{\kp^2\over m\beta^3} \sum_{\om_{1}\;\om_2\;\nu}
\int \ddp\ddk\ddq\;\G (\om_{1},\p)\;e^{i\om_1\eta}\;
\G (\om_{2},\k)\; e^{i\om_2\eta}\cr
&\qquad\qquad\qquad\times\G (\om_{2}+\nu,\k+\q)
{ (\p -\k )\cdot\q \over (\p -\k )^2\;\q ^2} \cr
&\cr
&=-\zeta{\kp^2\over 2m}\int\ddp\ddk\ddq\;\bigl( n_\p n_{\k +\q} +2\zeta n_\k
n_\p
 n_{\k +\q}\bigr)
{ (\p -\k )\cdot\q \over (\p -\k )^2\;\q ^2}\, . &(14) \cr
}
$$
{}From these equations it is straightforward to notice the following relation
$$
P^b=-(1+\zeta)(P^e +P^f)\, ,\eqno (15)
$$
reflecting the cancellation of exchange diagrams in the bosonic case.
Identity (15) implies a substantial reduction of calculations; to find $P^b
+P^e +P^f$, the only integral to compute is (12).

There are two different terms in (12): one involving the product of two $n$'s
and another with three $n$'s. They give quite distinct contributions:
$$
\eqalignno{
P^b +P^e +P^f=& {\kp^2\over 4\pi m\lt^4}\ln ^2(1-\zeta z)\int_0^{\x} dx\;\Phi
(x) \cr
 +{\kp^2\over 2\pi m\lt^4}&\sum_{l\geq 3}(\zeta
z)^l\sum_{n=1}^{l-2}\sum_{s=1}^{l-n-1}\;{1\over s(l-n-s)}\int_0^{\x} dx\;\Phi
(x)\; e^{-{ln\over s(l-n-s)}x^2}\, . &(16) \cr
}
$$
Here $\x$ is a UV cutoff and $\Phi (x)$ is the plasma dispersion function
[14],
$$
\Phi(x) = 2 \int_0^x dt\,{e^{-t^2}\,t\over\sqrt{x^2 - t^2}} =
2 e^{-x^2} \int_0^x dt\, e^{t^2}\, ,\eqno(17)
$$
its derivative being given by
$$
\Phi '(x)=2-2x\Phi (x)\;.
\eqno(18)
$$
The UV divergence comes from the loop with the seagull vertex and
from the divergence in the second Born approximation for scattering with a
$\delta$-function interaction [15]. These divergences will cancel when we
consider the whole set of diagrams.

We evaluate now the ring contribution from graph (d) of fig. 1,
$$
P^{d} = {\kp^2\over 2\beta} \sum_{\nu}
\int {\ddq}\,{\Pi^{0}{}_{00}(\nu,\q)\,
\Xi^{0}(\nu,\q)\over \q ^2}\, . \eqno(19)
$$
where $\Pi^{0}{}_{00}(\nu,\q)$ and $\Xi^{0}(\nu,\q )$ represent the
density-density correlator and the transverse component of the
current-current correlator respectively at the lowest order,
$$ \eqalignno{ \Pi^0{}_{ij}(\nu,\q) &=
-{\zeta\over m^2}{1\over\beta}\sum_{\omega_1} \int \ddp\, \Bigl[(\p +
{\q\over 2})_i\,(\p +{\q\over 2})_j\, \G
(\omega_{1},{\bf p})\, \G (\omega_{1}+\nu,\q + \p ) \cr
&\qquad+ m\,\delta_{ij}\,e^{i\omega_1 \eta}\,\G (\omega_{1},\p )\,\Bigr]\cr
&= -\Pi^0{}_{00}\,(\nu_{n},{\bf q})\,{\nu_n^2\over{\bf q}^2}\,
     {q_i\, q_j\over{\bf q}^2} +
    \Xi^{0}\,(\nu_{n},{\bf q})\,(\delta_{ij} - {q_i\, q_j\over{\bf q}^2})\,.
     &(20)\cr }
$$
Performing the sums and integrals, it yields (the first part comes from
two-$n$ terms; the second, from three-$n$ terms)
$$
\eqalignno{
P^d=& {\kp^2\over 8\pi m\lt^4}\ln ^2(1-\zeta z)\int_0^{\x} dx\;\biggl({\Phi
(x)\over x^2}-{2\over x}\biggr) \cr
& +{\kp^2\over 4\pi m\lt^4}\sum_{l\geq 3}(\zeta
z)^l\sum_{n=1}^{l-2}\sum_{s=1}^{l-n-1}\;{l-n-s\over s(l-s)^2}\int_0^{\x}
dx\;\biggl({\Phi
(x)\over x^2}-{2\over x}\biggr)\; e^{-{ln\over s(l-n-s)}x^2}\, . &(21) \cr
}
$$
The ring loop with a seagull vertex has yielded a UV logarithmic
divergence. This exactly cancels the infinities we have found before. To
see how this happens, we quote some results about the plasma dispersion
function, namely
$$
\int_0^\infty dx\; e^{-ax^2}\;\biggl({\Phi (x)\over x^2}-{2\over x}+2(1+a)\Phi
(x)\biggr)=2\,,
\eqno(22)
$$
$$
\int_0^\infty dx\; e^{-ax^2}\;\Phi (x) =
{1\over 2\sqrt{1+a}}\ln \biggl({\sqrt{1+a}+1\over\sqrt{1+a}-1}\biggr) \,.
\eqno(23)
$$
Both of them follow from eqs. (17) and (18). Now, first use eq. (22) (with
$a=0$) to compute the two-$n$ contribution from eqs. (16) and (19):
$$
{\kp^2\over 4\pi m\lt^4}\ln ^2(1-\zeta z)\, . \eqno (24)
$$
Split the three-$n$ term in eq. (16) in two parts: one is what must be added to
the three-$n$ term in eq. (21) so that eq. (22) can be used. The sum in $z$ can
be performed:
$$
-{\kp^2\over 4\pi m\lt^4}\ln (1-\zeta z)\;\biggl( {\zeta z\over 1-\zeta
z}+\ln(1-\zeta z)\biggr)\, . \eqno (25)
$$
The remaining part of eq. (16) is integrated using eq. (23):
$$
{\kp^2\over 4\pi m\lt^4}\sum_{l\geq 4}(\zeta
z)^l\sum_{n=1}^{l-2}\sum_{s=1}^{l-n-1}\;
{\alpha_{lns}\over\beta_{lns}}
\ln\biggl({\beta_{lns}+1\over\beta_{lns}-1}\biggr)
\, ,\eqno (26)
$$
where
$$
\alpha_{lns}\equiv{n(2s-l+n)\over s^2(l-s)(l-n-s)} \qquad , \quad
\beta_{lns}\equiv\sqrt{(l-s)(n+s)\over s(l-n-s)}\,.
\eqno (27)
$$
Due to the logarithms, it does not seem possible to perform the sums.
Notice also
that these terms give no correction to the first three cluster (and virial)
coefficients.

It only remains to evaluate diagrams (c), (g), (h). These are the
easiest to compute. They yield finite results:
$$
\eqalignno{
P^c=&{1+\zeta\over 2}{\kp^2\over \pi m\lt^4}{z\over 1-\zeta z}\ln ^2(1-\zeta
z)\, , &(28)\cr
P^g=&-{\kp^2\over 48\pi m\lt^4}{z\over 1-\zeta z}\ln ^2(1-\zeta
z)\, , &(29)\cr
P^h=&{\kp^2\over 4\pi m\lt^4}\ln (1-\zeta z)\;\biggl[ {\zeta z\over 1-\zeta
z}+\ln(1-\zeta z)\biggr]\, . &(30)\cr
}
$$

The solution we have found consists of a summable part (eqs. (24), (25), (28),
(29) and (30)) and a non-summable logarithmic part, eq. (26).
Collecting the summable terms we get
$$
{\kp^2\over 2\pi m\lt^4}\;\biggl( {1\over 2}+(\zeta +1-{1\over
24}){z\over 1-\zeta z}\biggr)\ln ^2(1-\zeta z)\, . \eqno (31)
$$
We find complete agreement with the results of ref. [11]. The third virial
coefficient obtained from these results does not depend on $\zeta$;
hence it satisfies  the `mirror symmetry' discovered by Sen [3]. For higher
virial coefficients, this symmetry does not hold.

To conclude, we have shown that the model based on standard perturbative
Chern-Simons field theory coupled to non-relativistic matter accounts for the
thermodynamic properties of the anyon gas. As an illustration, we have derived
a closed formula for the n-cluster coefficient through second order in the
anyon coupling constant.

\vskip 1.5 cm

Financial support under contract AEN90-0330 and UPV172.310-E035/90 is
acknowledged. R.E. acknowledges the Ministerio de Educaci\'on y Ciencia for
an FPI grant.

\vfill
\eject
\beginsection References

\frenchspacing
\item{[1]} For a recent review of this subject see F. Wilczek, Fractional
statistics and anyon superconductivity (World Scientific, Singapore, 1990);
S. Forte, Rev. Mod. Phys. {\bf64} (1992) 193.

\item{[2]} D. P. Arovas, R. Schrieffer, F. Wilczek and A. Zee, Nucl.
Phys. B {\bf 251} (1985) 117; see also D. P. Arovas, in: Geometric Phases
in Physics, eds. A. Shapere and F. Wilczek (World Scientific, Singapore,1989)
p.284.

\item{[3]} D. Sen, Phys. Rev. Lett. {\bf68} (1992) 2977.

\item{[4]} D. Sen, Phys. Rev. D {\bf46} (1992) 1846.

\item{[5]} J. Law, A. Suzuki and R.K. Bhadury, Phys. Rev. A {\bf46} (1992)
4693.

\item{[6]} M. Sporre, J.J.M. Verbaarschot and I. Zahed, Nucl. Phys.
B {\bf389} (1993) 645.

\item{[7]} J. Myrheim and K. Olaussen, Phys. Lett. B {\bf299} (1993) 267.

\item{[8]} J. McCabe and S. Ouvry, Phys. Lett. B {\bf260} (1991) 113.

\item{[9]} A. Comtet, J. McCabe and S. Ouvry, Phys. Lett. B {\bf260} (1991)
372.

\item{[10]} A. Dasni\`eres and S. Ouvry, Phys. Lett. B {\bf291} (1992) 130.

\item{[11]} A. Dasni\`eres and S. Ouvry, Nucl. Phys. B {\bf388} (1992) 715.

\item{[12]} M.A. Valle, Pressure in Chern-Simons field theory to three-loop
order, preprint EHU-FT-92/3 (December 1992), Phys. Lett. B, to appear.

\item{[13]} J. I. Kapusta, Finite-Temperature Field Theory (Cambridge
University Press, Cambridge, 1990).

\item{[14]} A. L. Fetter and J. D. Walecka, Quantum Theory of Many-Particle
Systems (McGraw-Hill, New York, 1971), sec. 33.

\item{[15]} R. Jackiw, Delta-function potentials in two
 and three-dimensional quantum mechanics, MIT preprint CPT 1937
(January 1991); C. Manuel and R. Tarrach, Phys. Lett. B {\bf268} (1991) 222.

\vfill
\eject

\beginsection Figure Captions

Fig. 1. Non zero diagrams contributing to the pressure to the second order in
$\kappa$. The sign $\pm$ refers to Bose or Fermi propagators. Graphs (c), (g)
and (h) do not contribute to the second virial coefficient. Combinatoric
factors are shown in the diagram.

\vfill
\eject

\end


\def\doublespace{\baselineskip=20pt plus 2pt
\lineskip=3pt minus 1pt\lineskiplimit=2pt}

\def\nofirstpagenoten{\nopagenumbers\footline={\ifnum\pageno>1\tenrm
\hss\folio\hss\fi}}
\def\nofirstpagenotwelve{\nopagenumbers\footline={\ifnum\pageno>1\twelverm
\hss\folio\hss\fi}}
\def\leaderfill{\leaders\hbox to 1em{\hss.\hss}\hfill}


\parindent=20pt
\def\narrow{\advance\leftskip by 40pt \advance\rightskip by 40pt}

\def\nonarrower{\advance\leftskip by -40pt\advance\rightskip by -40pt}

\def\boxit#1{\vbox{\hrule\hbox{\vrule\kern3pt
        \vbox{\kern3pt#1\kern3pt}\kern3pt\vrule}\hrule}}

\def\gtorder{\mathrel{\raise.3ex\hbox{$>$}\mkern-14mu
             \lower0.6ex\hbox{$\sim$}}}
\def\ltorder{\mathrel{\raise.3ex\hbox{$<$}|mkern-14mu
             \lower0.6ex\hbox{\sim$}}}
\def\dalemb#1#2{{\vbox{\hrule height .#2pt
        \hbox{\vrule width.#2pt height#1pt \kern#1pt
                \vrule width.#2pt}
        \hrule height.#2pt}}}

\font\twelvett=cmtt12 \font\twelvebf=cmbx12
\font\twelverm=cmr12 \font\twelvei=cmmi12 \font\twelvess=cmss12
\font\twelvesy=cmsy10 scaled \magstep1 \font\twelvesl=cmsl12
\font\twelveex=cmex10 scaled \magstep1 \font\twelveit=cmti12
\font\tenss=cmss10
 
 \font\ninebf=cmbx9
\font\ninerm=cmr9 \font\ninei=cmmi9
\font\ninesy=cmsy9 
\font\eightrm=cmr8
\catcode`@=11
\newskip\ttglue
\newfam\ssfam

\def\twelvepoint{\def\rm{\fam0\twelverm}
\textfont0=\twelverm \scriptfont0=\ninerm \scriptscriptfont0=\sevenrm
\textfont1=\twelvei \scriptfont1=\ninei \scriptscriptfont1=\seveni
\textfont2=\twelvesy \scriptfont2=\ninesy \scriptscriptfont2=\sevensy
\textfont3=\twelveex \scriptfont3=\twelveex \scriptscriptfont3=\twelveex
\def\it{\fam\itfam\twelveit} \textfont\itfam=\twelveit
\def\sl{\fam\slfam\twelvesl} \textfont\slfam=\twelvesl
\def\bf{\fam\bffam\twelvebf} \textfont\bffam=\twelvebf
\scriptfont\bffam=\ninebf \scriptscriptfont\bffam=\sevenbf
\def\tt{\fam\ttfam\twelvett} \textfont\ttfam=\twelvett
\def\ss{\fam\ssfam\twelvess} \textfont\ssfam=\twelvess
\tt \ttglue=.5em plus .25em minus .15em
\normalbaselineskip=14pt
\abovedisplayskip=14pt plus 3pt minus 10pt
\belowdisplayskip=14pt plus 3pt minus 10pt
\abovedisplayshortskip=0pt plus 3pt
\belowdisplayshortskip=8pt plus 3pt minus 5pt
\parskip=3pt plus 1.5pt
\setbox\strutbox=\hbox{\vrule height10pt depth4pt width0pt}
\let\sc=\ninerm
\let\big=\twelvebig \normalbaselines\rm}
\def\twelvebig#1{{\hbox{$\left#1\vbox to10pt{}\right.\n@space$}}}

\def\tenpoint{\def\rm{\fam0\tenrm}
\textfont0=\tenrm \scriptfont0=\sevenrm \scriptscriptfont0=\fiverm
\textfont1=\teni \scriptfont1=\seveni \scriptscriptfont1=\fivei
\textfont2=\tensy \scriptfont2=\sevensy \scriptscriptfont2=\fivesy
\textfont3=\tenex \scriptfont3=\tenex \scriptscriptfont3=\tenex
\def\it{\fam\itfam\tenit} \textfont\itfam=\tenit
\def\sl{\fam\slfam\tensl} \textfont\slfam=\tensl
\def\bf{\fam\bffam\tenbf} \textfont\bffam=\tenbf
\scriptfont\bffam=\sevenbf \scriptscriptfont\bffam=\fivebf
\def\tt{\fam\ttfam\tentt} \textfont\ttfam=\tentt
\def\ss{\fam\ssfam\tenss} \textfont\ssfam=\tenss
\tt \ttglue=.5em plus .25em minus .15em
\normalbaselineskip=12pt
\abovedisplayskip=12pt plus 3pt minus 9pt
\belowdisplayskip=12pt plus 3pt minus 9pt
\abovedisplayshortskip=0pt plus 3pt
\belowdisplayshortskip=7pt plus 3pt minus 4pt
\parskip=0.0pt plus 1.0pt
\setbox\strutbox=\hbox{\vrule height8.5pt depth3.5pt width0pt}
\let\sc=\eightrm
\let\big=\tenbig \normalbaselines\rm}
\def\tenbig#1{{\hbox{$\left#1\vbox to8.5pt{}\right.\n@space$}}}
\let\rawfootnote=\footnote \def\footnote#1#2{{\rm\parskip=0pt\rawfootnote{#1}
{#2\hfill\vrule height 0pt depth 6pt width 0pt}}}